\documentclass[twocolumn,pra,aps,showpacs]{revtex4}
\usepackage{amsmath}
\usepackage{color}
\usepackage{graphicx}

\begin{document}

\title{ Microwave transmission through an artificial atomic chain coupled to a superconducting photonic crystal }

\author{Guo-Zhu Song$^{1}$, Leong-Chuan Kwek$^{2,3,4,5}$, Fu-Guo Deng$^{6,7}$, and Gui-Lu Long$^{1,8,9}$\footnote{Corresponding author: gllong@tsinghua.edu.cn } }

\address{$^{1}$State Key Laboratory of Low-Dimensional Quantum Physics and Department of Physics, Tsinghua University, Beijing 100084, China\\
$^{2}$Centre for Quantum Technologies, National University of Singapore, 3 Science Drive 2, Singapore 117543\\
$^{3}$Institute of Advanced Studies, Nanyang Technological University, Singapore 639673\\
$^{4}$National Institute of Education, Nanyang Technological University, Singapore 637616\\
$^{5}$MajuLab, CNRS-UNS-NUS-NTU International Joint Research Unit, UMI 3654, Singapore\\
$^{6}$Department of Physics, Applied Optics Beijing Area Major Laboratory, Beijing Normal University, Beijing 100875, China\\
$^{7}$NAAM-Research Group, Department of Mathematics, Faculty of Science, King Abdulaziz University, P.O. Box 80203, Jeddah 21589, Saudi Arabia\\
$^{8}$Tsinghua National Laboratory of Information Science and Technology, Beijing 100084, China\\
$^{9}$Collaborative Innovation Center of Quantum Matter, Beijing 100084, China  }
\date{\today }

\begin{abstract}
Emitters strongly coupled to a photonic crystal provide a powerful platform for realizing novel quantum light-matter interactions. Here we study the optical properties of a three-level artificial
atomic chain coupled to a one-dimensional superconducting microwave photonic crystal. A sharp minimum-energy dip appears in the transmission spectrum of a weak input field, which reveals rich behavior of the long-range interactions arising from localized bound states. We find that the dip frequency
scales linearly with both the number of the artificial atoms and the characteristic strength of the long-range interactions when the localization length of the bound state is sufficiently large. Motivated by this observation, we present a simple model to calculate the dip frequency with system parameters, which agrees well with the results from exact numerics for large localization lengths. We observe oscillation between bunching and antibunching in photon-photon correlation function of the output field. Furthermore, we find
that the model remains valid even though the coupling strengths between the photonic crystal and artificial atoms are not exactly equal and the phases of external driving fields for the artificial atoms are different.
Thus, we may infer valuable system parameters from the dip location in the transmission spectrum, which
provides an important measuring tool for the superconducting microwave photonic crystal systems in experiment. With remarkable advances to couple artificial atoms with microwave photonic crystals, our proposal may be experimentally realized in currently available superconducting circuits.
\end{abstract}
\pacs{03.67.Lx, 03.67.Pp, 42.50.Ex, 42.50.Pq}

\maketitle

\section{Introduction}

In recent years, one-dimensional (1D) waveguide quantum electrodynamics (QED) systems have emerged
as an exciting frontier in quantum information science \cite{ShenOL2005,Shen2007PRL,Shen2009pra,Fan2010pra,Zhou2013prl,PLodahl2015rmp,Liao2016,Diby2017rmp,Das2017PRL,Manzoni2017,Xia2018PRL,DEChang2018,Shen2018pra}.
The waveguide systems benefit from the confinement of continuous electromagnetic modes over
a large bandwidth, which couple to nearby atoms or embedded artificial atoms. There are a
wide variety of systems that can act as waveguide platforms such as plasmonic nanowires
\cite{AkimovNature2007,Chang2007nap,Tudela2011prl,Akselrod2014}, optical nanofibers \cite{DayanScience2008,Vetsch2010prl,RausPRL2011,DReitz2013PRL,Petersen2014,Kien2014,Mitsch2014,GoutaudPRL2015,CSayrin2015,HLsorensen2016,Song2017AOP,PSolano2017,Cheng2017pra,song2017pra,Cheng2018PRA,Yanwb2018,Yan2018}, diamond waveguides \cite{BabinecNat2010,ClaudonPhoton2010,Clevenson2015,Sipahigil2016},
superconducting transmission lines \cite{WallraffNature2004,ShenPRL2005,Lzhou2008,AstafievScience2010,LooSci2013,klalum2013,GuPR2017,Kockum2018,MMirho2018,PMHarr2018},
and photonic crystal waveguides \cite{JDJoan2008book}. Due to the
intrinsically tailorable band structure, photonic crystal waveguides are
of particular interest and enable tunable long-range interactions in many-body systems
\cite{TudelaNAT2015,JSDouglas2015,jsDoug2016prx,EWAN2017,Cirac2018pra,Song2018}.

Photonic crystals are highly dispersive periodic dielectric media in which the
refractive index is modulated periodically due to photonic band gaps \cite{JDJoan2008book,GanL2015}.
In this configuration, when a qubit trapped nearby the photonic crystal is excited
at a frequency inside the band gap, it cannot radiate into the dielectric but gives
rise to a qubit-photon bound state \cite{SJohn1990prl,SJohn1990prb,SJohn1995prl,Rabl2016pra,ShiT2016,Bello2018}.
The photonic component of the bound state is an exponentially decaying envelope
spatially centered at the qubit position, which facilitates coherent excitation
exchange with proximal qubits \cite{JSDouglas2015,jsDoug2016prx}. Although
significant progress has been made \cite{TLHansen2008,ALaucht2012,JDThompson2013,TGTiecke2014,MArcari2014prl,SPYuapl2014,Tudela2015prl,ABYoung2015,AGoban2015PRL,Sollner2015,Lefeber2015,AGobannatc2014,JDHood2016PNAS,Litao2018,Yu2018arxiv,Tudela2018arxiv,Burgersa2019}, realizing efficient coupling in the optical regime faces the challenging task of
interfacing emitters with photonic crystal waveguides. Recently, superconducting quantum
circuits provide an alternative approach to study the physics of the bound state
in the microwave regime \cite{GuPR2017}. Using a stepped-impedance microwave photonic crystal
and a superconducting transmon qubit, Liu \emph{et al}. first experimentally observed
the bound state in superconducting transmission lines \cite{YLiu2017}. In their device,
by adjusting the detuning between the qubit and the band edge, the localization length
of the bound state is widely tunable. Later, Sundaresan \emph{et al}. experimentally
realize strong coupling between two transmon qubits and a superconducting microwave
photonic crystal, which is a promising benchmark to create 1D chains of qubit-photon
bound states with tunable long-range interactions \cite{NMSun2018}.

Inspired by these remarkable advances, we here study the scattering properties of a weak
incident field travelling through an array of $\Delta$-type artificial atoms
coupled to a superconducting microwave photonic crystal. In this work, one transition
of the $\Delta$-type artificial atom is inside the band gap, which gives rise to the
bound state. Besides, another transition is coupled to the electromagnetic modes of the
superconducting microwave photonic crystal, and is utilized to explore the long-range
interactions arising from the bound states.

With an effective non-Hermitian Hamiltonian,
we calculate the transmission spectrum of a weak microwave input field and observe a sharp
minimum-energy dip. We analyze the relation between the dip frequency and the system
parameters, such as the number of the artificial atoms and the characteristic strength
of the long-range interactions. The results reveal that the dip frequency scales linearly
with both the number of the artificial atoms and the characteristic strength for large
localization lengths of the bound states, which may provide an important measuring tool for the superconducting
microwave photonic crystal systems. Motivated by this observation, we give a simple model to calculate the dip frequency with system parameters. We find that,
when the localization length of the bound state is large enough, the results of our model agree well with
exact numerics. We also study the photon-photon correlation function in the resonant case, and observe oscillation between bunching and antibunching for both transmitted and reflected fields. Moreover, we study the effects of a Gaussian inhomogeneous broadening
of the coupling strength and different phases of the external driving field for the artificial
atoms on the dip frequency, respectively. The results show that, our model remains valid even
though the coupling strengths between the photonic crystal and artificial atoms are not
exactly equal and the phases of the driving fields for the artificial atoms are different.
That is, in experiment, one may infer the system parameters by measuring the dip frequency
in the transmission spectrum.

This article is organized as follows: In Sec. \ref{MODEL}, we present the physics
of an artificial atomic chain coupled to a superconducting microwave photonic
crystal, and an effective Hamiltonian is introduced for the system. In Sec. \ref{RESULTS},
we study the transmission spectrum of the weak microwave coherent field, and give
a simple model to estimate the dip frequency with system parameters. We also calculate the photon-photon
correlation function of the output field in the resonant case.
Moreover, we analyze the effects of the inhomogeneous broadening of the coupling
strength, different phases of the driving field for the artificial atoms, and the dissipation of the bound states on the
dip frequency, respectively. In Sec. \ref{expermental}, we discuss the feasibility of our system with
recent experiments. Finally, we summarize the results, and discuss the
advantage of the superconducting microwave photonic crystal systems
in Sec. \ref{discussion}.

\section{MODEL AND HAMILTONIAN}  \label{MODEL}

\begin{figure}[htbp]      
\centering\includegraphics[width=8.1cm]{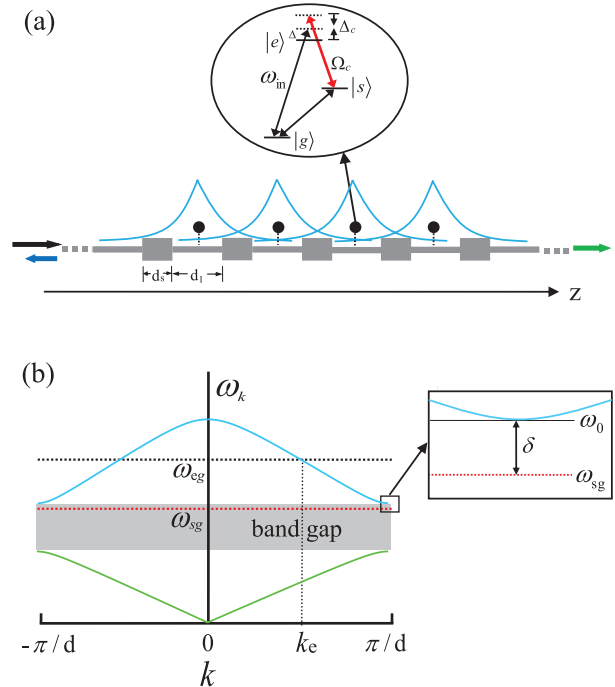} \caption{(a) Schematic diagram of the microwave
transport through a three-level artificial atomic chain (black dots) coupled to an infinite superconducting
microwave photonic crystal (gray line of alternating width) with unit cell length $d=d_{l}+d_{s}$.
A coherent microwave field (black arrow) is injected from left to interact with the artificial
atomic chain, and then a reflected field (blue arrow) and a transmitted field (green arrow) are
collected at the left and right exports, respectively. The blue envelopes represent the photonic
components of the bound states, which are spatially centered at the locations of the artificial atoms.
(b) Band structure of the 1D superconducting microwave photonic crystal.
The green and blue solid lines represent the first and second photonic bands,
respectively. The resonance frequency $\omega_{sg}$ (horizontal red dashed
line) is inside the band gap (gray region), and is close to the second band edge
frequency $\omega_{_{0}}$ with detuning $\delta=\omega_{_{0}}-\omega_{sg}$. The
resonance frequency $\omega_{eg}$ (horizontal black dashed line) lies in the
second band, and the corresponding wave vector is $k_{e}$.
}
\label{figure1}
\end{figure}

We model an array of $\Delta$-type artificial atoms coupled to an infinite superconducting
microwave photonic crystal, as shown in Fig. \ref{figure1}(a).
The superconducting microwave photonic crystal is implemented by periodically alternating
sections of low and high impedance coplanar waveguides, which can be realized via changing
the gap width and centre pin of the coplanar waveguide \cite{YLiu2017,NMSun2018}. The dispersion
relation of the guided modes in the superconducting microwave photonic crystal can
be obtained by the transfer matrix method \cite{YLiu2017}, and is given by
\begin{eqnarray}     \label{eqa1}       
\begin{split}
\cos(kd)\!=\!\cos(\frac{\omega_{_{k}} d}{v_{g}})\!\!-\!\!\frac{1}{2}(\!\frac{Z_{l}}{Z_{s}}\!\!+\!\!\frac{Z_{s}}{Z_{l}}\!\!-\!\!2)\!\sin(\frac{\omega_{_{k}} d_{l}}{v_{g}})\!\sin(\frac{\omega_{_{k}} d_{s}}{v_{g}}),
\end{split}
\end{eqnarray}
where $k$ is Bloch wave vector, $d=d_{l}+d_{s}$ is the unit cell length, and $\omega_{_{k}}$ is the guided mode frequency with phase velocity $v_{g}$. $Z_{l}$ ($Z_{s}$) and $d_{l}$ ($d_{s}$) represent the impedance and length of the high (low) impedance coplanar waveguide in the unit cell, respectively. Each artificial atom has three energy levels $|g\rangle$, $|e\rangle$ and $|s\rangle$.  The $\Delta$-type artificial atom can be realized by a flux-based superconducting quantum circuit when the external magnetic flux through the loop $\Phi_{e}\neq\Phi_{_{0}}/2$, where $\Phi_{_{0}}$ is the flux quantum \cite{Liu2005,Jia2017}. Since the second band is smoother than the first one,
each artificial atom is purposely placed in the center of the high-impedance section, which maximizes
(minimizes) the coupling between the artificial atom and the second (first) band \cite{YLiu2017,NMSun2018}.
Moreover, the width of the second band can be sufficiently large so that the influence of other band is ignored.
In detail, the $\Delta$-type artificial atom is coupled to the high-impedance section of the unit cell
through the loop-line mutual inductance $M$.

Here, we assume that the resonance frequency
$\omega_{sg}$ of the transition $|g\rangle\leftrightarrow|s\rangle$ is inside the band gap, and is
close to second band edge frequency $\omega_{_{0}}$ with detuning $\delta\!=\!\omega_{_{0}}\!-\!\omega_{sg}$, as shown in Fig. \ref{figure1}(b). In this domain, due to the van Hove singularity in the density of states, the transition $|g\rangle\leftrightarrow|s\rangle$ of the artificial atom is predominantly coupled
to the modes close to the second band edge. In this case, we can approximate the dispersion relation
near the second band edge to be quadratic $\omega_{_{k}}\approx\omega_{_{0}}+\alpha d^{2}(k-k_{_{0}})^2$,
where $\alpha$ denotes the curvature of the band structure and $k_{_{0}}=\pi/d$ is the band edge wave vector.
Once such an artificial atom is excited to the state $|s\rangle$, a localized bound state appears \cite{JSDouglas2015}. Specifically, as
shown in Fig. \ref{figure1}(a), in real space, the photonic component of the bound state
is exponentially localized around the artificial atom with localization length $L$, which mediates
long-range coherent dipole-dipole interactions with proximal artificial atoms. Since $L\propto\sqrt{1/\delta}$,
we can tune the localization length by altering the detuning between the resonance frequency $\omega_{sg}$
and the second band edge frequency $\omega_{_{0}}$.

While it is not allowed to probe this coherent interactions using the input field
around frequency $\omega_{sg}$ due to the existence of the band gap. Thus, we assume that the resonance
frequency $\omega_{eg}$ of the transition $|g\rangle\leftrightarrow|e\rangle$ lies in the second band,
which gives rise to the coupling between the transition $|g\rangle\leftrightarrow|e\rangle$ and the modes
of the second band. Here, the coupling strengths between all artificial atoms and the second
band are assumed to be identical. The single-atom coupling strength to the second band is given by $\text{g}$, where
$\text{g}\!\propto \!MI_{p}$ with $I_{p}$ being the amplitude of the persistent current in the loop.
That is, we can use the second photonic band as a conventional superconducting transmission line to collect the
scattered fields and explore the long-range interactions between the artificial atoms. In addition,
with a magnetic flux coil, we introduce an external microwave field
(Rabi frequency $\Omega_{c}$) to drive the transition $|e\rangle\leftrightarrow|s\rangle$ of each artificial atom.
In fact, the transition $|e\rangle\leftrightarrow|s\rangle$ driven by an external field connects the two mechanisms mentioned above, i.e., the long-range interactions produced by the transition $|g\rangle\leftrightarrow|s\rangle$ and the detecting channel arising from the coupling between the transition $|g\rangle\leftrightarrow|e\rangle$ and the second band.

The system composed of the artificial atomic chain and the superconducting microwave photonic crystal
can be described by an effective non-Hermitian Hamiltonian \cite{Chang2012,JSDouglas2015,JDHood2016PNAS,NMSun2018}
\begin{eqnarray}     \label{eqa2}       
H_{non}=\!\!\!&&-{{\sum\limits_{j}^n}}\big[(\Delta\!+\!i\Gamma_{e}^{'}/2)\sigma_{ee}^{j}\!+\!(\Delta-\Delta_{c}+\!i\Gamma_{s}^{'}/2) \sigma_{ss}^{j}\nonumber\\
&&+\Omega_{c}(\sigma_{es}^{j}+\text{H.c.}\big]\!-\!i\frac{\Gamma_{_{0}}}{2}{{\sum\limits_{j,k}^n}}e^{ik_{e}|z_{_{j}}-z_{_{k}}|}\sigma_{eg}^{j}\sigma_{ge}^{k}\nonumber\\
&&-V{{\sum\limits_{j, k}^n}}(-1)^{|z_{_{j}}-z_{_{k}}|/d}e^{-|z_{_{j}}-z_{_{k}}|/L}\sigma_{sg}^{j}\sigma_{gs}^{k}.
\end{eqnarray}
Here $\Delta\!=\!\omega_{in}-\omega_{eg}$ is the detuning between the frequency $\omega_{in}$ of
the input field with wave vector $k_{in}$ and the resonance frequency $\omega_{eg}$. $\Gamma_{e}^{'}$
($\Gamma_{s}^{'}$) is the decay rate of the state $|e\rangle$ ($|s\rangle$) into free
space, $n$ is the number of the artificial atoms, and $\Delta_{c}\!=\!\omega_{c}-\omega_{es}$ denotes the
detuning between the frequency $\omega_{c}$ of the external driving field and the frequency $\omega_{es}$
of the transition $|e\rangle\!\leftrightarrow\!|s\rangle$. $\Gamma_{_{0}}\!=\!4\pi\text{g}^2/v_{g}$ is
the single-atom spontaneous emission rate into the photonic crystal modes. $z_{_{j}}$ represents the
position of artificial atom $j$, and $V$ characterizes the strength of the long-range coherent
interactions arising from the bound states. Note that, we include the case $j=k$ in the last term
of Eq. (\ref{eqa2}), which corresponds to the Stark shift experienced by artificial atom $j$ due to its own bound photon.

\begin{figure}
\centering\includegraphics[width = 7.9cm]{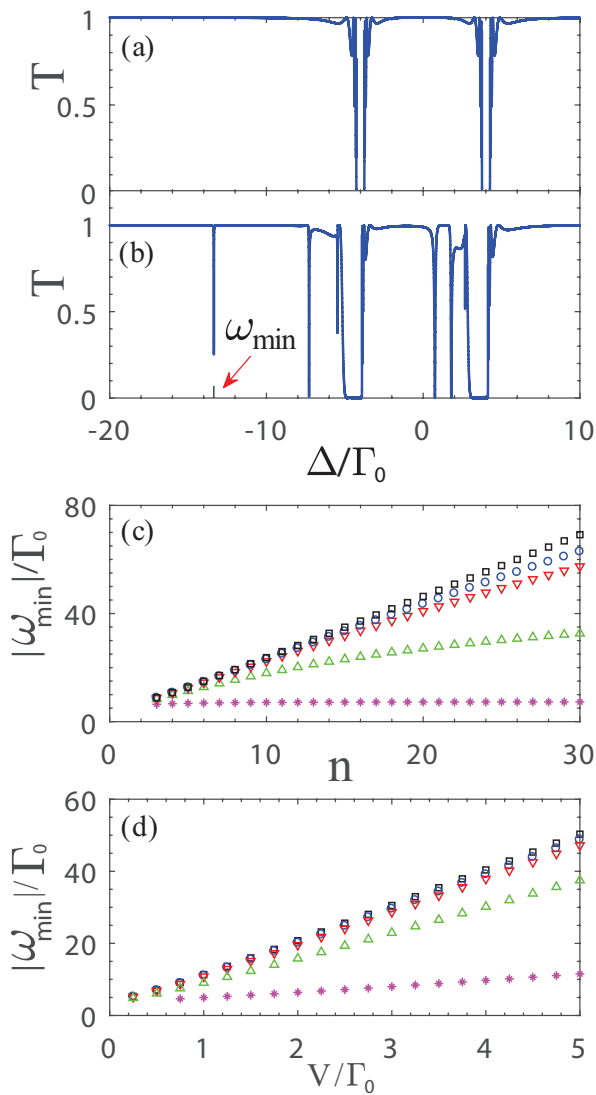} \caption{ The transmission spectra of the input
field as a function of the frequency detuning $\Delta/\Gamma_{0}$ for (a) $V=0$ and (b) $V=2.3\Gamma_{_{0}}$
with localization length $L\!=\!4.1d$ when $n=10$ artificial atoms are equally spaced along the
superconducting microwave photonic crystal. (c) The frequency $|\omega_{min}|$ versus the number $n$
of the artificial atoms for $L\!=\!10^{4}d$ (black squares), $L\!=\!10^{2}d$ (blue circles), $L\!=\!50d$
(red down-triangles), $L\!=\!10d$ (up green-triangles) and $L\!=\!d$ (purple asterisks) with $V=2.3\Gamma_{_{0}}$.
(d) The frequency $|\omega_{min}|$ as a function of the characteristic strength $V$ for $L\!=\!10^{4}d$ (black squares),
$L\!=\!10^{2}d$ (blue circles), $L\!=\!50d$ (red down-triangles), $L\!=\!10d$ (up green-triangles) and $L\!=\!d$
(purple asterisks) with $n=10$. Parameters: (a)-(d) $\mathcal {E}=0.0001\sqrt{\frac{\Gamma_{_{0}}}{2v_{g}}}$,
$k_{e}d\!=\!\pi/2$, $\Delta_{c}\!=\!0$, $\Omega_{c}\!=\!4\Gamma_{_{0}}$, $\Gamma_{e}'=9.5\times10^{-6}\Gamma_{_{0}}$,
and $\Gamma_{s}'=9.5\times10^{-6}\Gamma_{_{0}}$. } \label{figure2}
\end{figure}

Here, we mainly focus on the transport properties of a weak coherent probe field propagating
through the artificial atomic chain. The corresponding driving part is given by $H_{dr}\!=\!\sqrt{\frac{v_{g}\Gamma_{_{0}}}{2}}\mathcal {E}{{\sum\limits_{j}^n}}(\sigma_{eg}^{j}e^{ik_{in}z_{_{j}}}+\text{H.c.})$, where $\mathcal {E}$
is the amplitude of the weak probe field (Rabi frequency $\sqrt{\frac{v_{g}\Gamma_{_{0}}}{2}}\mathcal {E}$).
As a consequence, the dynamics of the system is governed by the total Hamiltonian $H=H_{non}+H_{dr}$,
and the initial state is the ground state $|\psi_{_{0}}\rangle=|g\rangle^{\otimes n}$ of the artificial atomic chain.
When the probe field is sufficiently weak ($\sqrt{\frac{v_{g}\Gamma_{_{0}}}{2}}\mathcal {E}\!\ll\!\Gamma_{e}^{'}$),
the occurrence of quantum jumps is infrequent and can be neglected \cite{Albrecht2017njp}. Using the input-output
formalism \cite{CanevaNJP2015}, the transmitted (T) and reflected (R) fields operators are
\begin{eqnarray}     \label{eqa3}       
a_{_{T}}(z) =&&\mathcal {E}e^{ik_{in}z}+i\sqrt{\frac{\Gamma_{_{0}}}{2v_{g}}}{{\sum\limits_{j}^n}}\sigma_{ge}^{j}e^{ik_{e}(z-z_{_{j}})},\nonumber\\
a_{_{R}}(z) =&&i\sqrt{\frac{\Gamma_{_{0}}}{2v_{g}}}{{\sum\limits_{j}^n}}\sigma_{ge}^{j}e^{-ik_{e}(z-z_{_{j}})}.
\end{eqnarray}
Consequently, the transmittance $T$ and reflection $R$ of the weak probe field are given by
\begin{eqnarray}     \label{eqa4}       
T=&&\frac{\langle\psi|a_{_{T}}^{\dagger}a_{_{T}}|\psi\rangle}{\mathcal {E}^{2}},\nonumber\\
R=&&\frac{\langle\psi|a_{_{R}}^{\dagger}a_{_{R}}|\psi\rangle}{\mathcal {E}^{2}},
\end{eqnarray}
where $|\psi\rangle$ represents the steady state.

\section{NUMERICAL RESULTS} \label{RESULTS}

\subsection{The transmission properties of the input field} \label{Transmission}

Here, we consider the case that $n\!=\!10$ artificial atoms are equally spaced along the
1D infinite superconducting microwave photonic crystal. To minimize the reflection of the
input field from the artificial atomic chain, we choose the configuration $k_{e}d\!=\!\pi/2$
\cite{Chang2011njp,CanevaNJP2015}. As shown in Figs. \ref{figure2}(a) and \ref{figure2}(b), we
calculate the transmission spectra of the weak monochromatic coherent input field with
localization length $L\!=\!4.1d$ for two choices of characteristic strength $V$. As
illustrated in Fig. \ref{figure2}(a), for $V\!=\!0$ (i.e., a conventional superconducting
transmission line), we observe the electromagnetically induced transparency phenomenon in
the transmission spectrum. While for $V\!\neq\!0$, such as for the case $V\!=\!2.3\Gamma_{_{0}}$
shown in Fig. \ref{figure2}(b), some new sharp dips emerge in the transmission spectrum,
which arise from the long-range coherent interactions between the artificial atoms.

In the following, we mainly focus on the minimum resonance frequency $\omega_{min}$, since
it may reveal rich behavior of the long-range interactions. For simplicity, we adopt
$|\omega_{min}|$ instead of $\omega_{min}$ in our discussions, which does not qualitatively
influence the conclusions. As shown in Fig. \ref{figure2}(c), we plot $|\omega_{min}|$ as
a function of the number $n$ of the artificial atoms coupled to the photonic crystal for different choices
of the localization length $L$. The results show that in the case $L/d\gg1$, $|\omega_{min}|$
scales linearly with the number of the artificial atoms. In particular, for infinite-range
interaction such as $L/d=10^{4}$, we conclude the relation $|\omega_{min}|\approx nV$. In
fact, by diagonalizing the last term of Eq. (\ref{eqa2}) for $L/d\rightarrow\infty$, we
obtain $(n-1)$ degenerate resonance energies zero and one resonance energy $-nV$. While,
for finite-range interaction such as $L/d=10$, the frequency $|\omega_{min}|$ scales sub-linearly
with the number of artificial atoms. Moreover, for short-range interaction such as $L\!=\!d$, the
frequency $|\omega_{min}|$ almost remains constant despite we increase the number of the
superconducting artificial atoms, as shown in Fig. \ref{figure2}(c). In addition, we also
give the frequency $|\omega_{min}|$ versus the characteristic strength $V$ for different choices
of the localization length $L$. Fig. \ref{figure2}(d) reveals that the frequency $|\omega_{min}|$
scales linearly with the characteristic strength $V$ in all cases.

\begin{figure}[tpb]    
\centering\includegraphics[width=8.0cm]{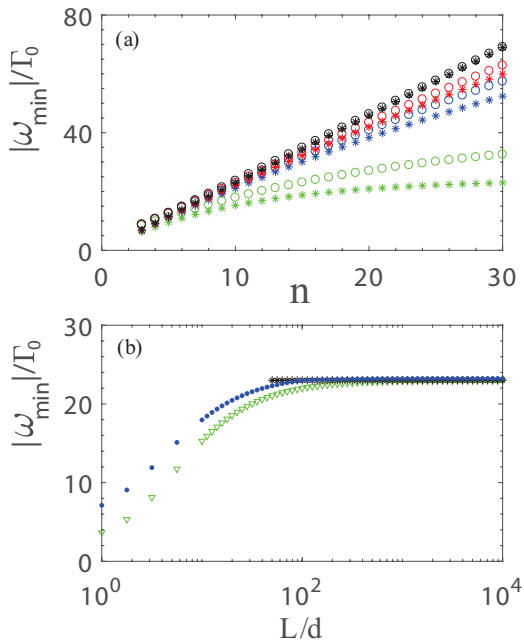} \caption{ (a) The frequency $|\omega_{min}|$
as a function of the number $n$ of the artificial atoms calculated from Eq. (\ref{eqa7})
(asterisks) and exact numerics (circles) for $L=10^{4}d$ (black), $L=10^{2}d$ (red), $L=50d$
(blue) and $L=10d$ (green).  (b)  The frequency $|\omega_{min}|$ versus the localization length
$L$ calculated from exact numerics (blue dots), Eq. (\ref{eqa7}) (green down-triangles), and
Eq. (\ref{eqa8}) (black asterisks) with $n=10$. Parameters: (a)-(b) $\mathcal {E}=0.0001\sqrt{\frac{\Gamma_{_{0}}}{2v_{g}}}$,
$V=2.3\Gamma_{_{0}}$, $k_{e}d\!=\!\pi/2$, $\Delta_{c}\!=\!0$, $\Omega_{c}\!=\!4\Gamma_{_{0}}$,
$\Gamma_{e}'=9.5\times10^{-6}\Gamma_{_{0}}$, and $\Gamma_{s}'=9.5\times10^{-6}\Gamma_{_{0}}$.
} \label{figure3}
\end{figure}

The above observation motivates us to present a simple model for estimating the frequency
$|\omega_{min}|$ with the parameters $V$, $L$, $n$. As shown in Figs. \ref{figure2}(a) and \ref{figure2}(b),
the minimum resonance energy $\omega_{min}$ arises from the long-range interaction Hamiltonian
\begin{eqnarray}     \label{eqa5}       
H_{b}\;=\;-V{{\sum\limits_{j, k}^n}}e^{-|z_{_{j}}-z_{_{k}}|/L}\sigma_{sg}^{j}\sigma_{gs}^{k}.
\end{eqnarray}
Here, for brevity, we have ignored the phase factor, which does not qualitatively change
the conclusions. Intuitively, for any two atoms $j$ and $k$ chosen from $n$ artificial
atoms, we may sum over all possible cases for separations, and get one energy $A$, i.e.,
\begin{eqnarray}     \label{eqa6}       
\begin{split}
A=-V\sum\limits_{l=0}^{n-1}e^{-ld/L}.
\end{split}
\end{eqnarray}
Note that, here we include the term $l=0$, which corresponds to Stark shift experienced
by an artificial atom due to its own bound photon. Summing all the terms, we easily obtain
\begin{eqnarray}     \label{eqa7}       
\begin{split}
|A|=V\frac{e^{d/L}-e^{-(n-1)d/L}}{e^{d/L}-1}.
\end{split}
\end{eqnarray}
The result of this model agrees well with the exact numerics shown in Fig. \ref{figure2}(d).
That is, the frequency $|\omega_{min}|$ calculated from exact numerics scales linearly with
the characteristic strength $V$ for fixed number of the artificial atoms and localization
length $L$ of the bound states. In the limit $L\gg nd$, using the approximation $e^{x}\approx(1+x)$
when $x\rightarrow0$, the energy $|A|$ in Eq. (\ref{eqa7}) can be written approximately as
\begin{eqnarray}     \label{eqa8}       
\begin{split}
|A|\approx nV. \;\;\;\;\;
\end{split}
\end{eqnarray}
It is consistent with the result $|\omega_{min}|\approx nV$ by diagonalizing the
long-range interaction Hamiltonian for infinite localization length mentioned above.
Consequently, with this model, in the limit $L\gg d$, we may use the energy $|A|$ to estimate the frequency $|\omega_{min}|$
with system parameters $L$, $V$, $n$ and $d$.

\begin{figure}[tpb]    
\centering\includegraphics[width=5.9cm]{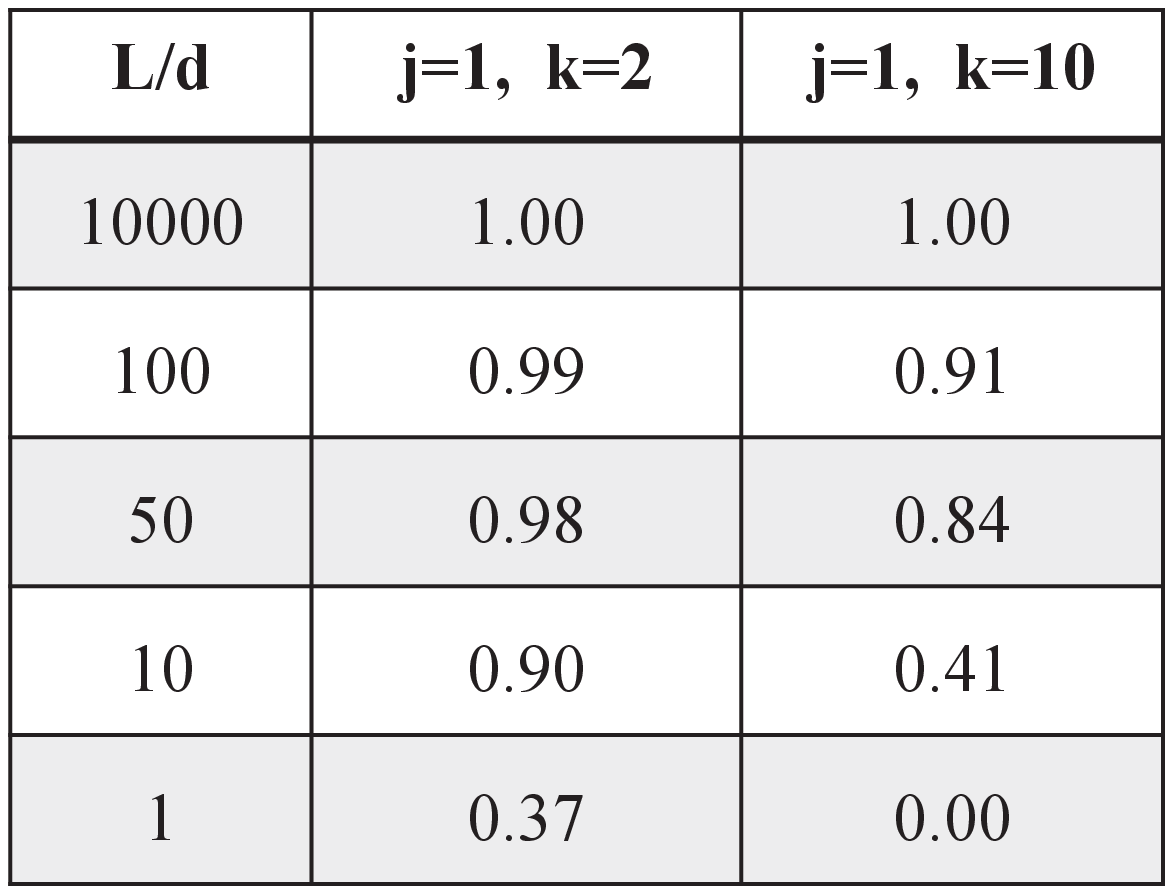} \caption{ The interaction strengths of the nearest two atoms ($j\!=\!1$, $k\!=\!2$)
and farthest two atoms ($j\!=\!1$, $k\!=\!10$) arising from Eq. (\ref{eqa5}) with $n\!=\!10$ for $L\!=\!10^{4}d$, $L\!=\!10^{2}d$, $L\!=\!50d$, $L\!=\!10d$ and $L\!=\!d$. The numbers (in units of $V$) in the second and third columns retain two digits after the decimal point. Here, we omit the minus sign in Eq. (\ref{eqa5}).  } \label{figure4}
\end{figure}

To demonstrate the validity of our model, we compare its results with exact numerics,
as shown in Fig. \ref{figure3}(a). We observe that this model agrees well with the exact
numerics in the limit $L\gg d$. While for finite localization length $L$, such as $L\!=\!10d$,
there is a difference between the two results, and the frequency $|\omega_{min}|$ scales
sub-linearly with the number $n$ of the artificial atoms in both cases. To obtain the validity of the
localization length $L$, we give the results from our model and exact numerics as a
function of the localization length $L$, respectively, as shown in Fig. \ref{figure3}(b).
We observe that, when $L\geq10^2d$ in our system, this simple model agrees well with
exact numerics. In other words, for a sufficiently large localization length $L$, one may
infer the characteristic strength $V$ of the long-range interactions from Eq. (\ref{eqa8})
for fixed number of the artificial atoms, by measuring the dip frequency $|\omega_{min}|$
in the transmission spectrum.

As shown in Fig. \ref{figure4}, with $n=10$ and five choices of the localization length $L$,
we give the coupling strengths of the nearest and farthest two artificial atoms arising from
the long-range coherent interactions in Eq. (\ref{eqa5}), respectively. For simplicity, here we
omit the minus sign in Eq. (\ref{eqa5}), which does not influence the conclusions. In fact, the
interaction strength between two atoms is determined by the overlap of their photonic envelopes
with the atoms, as shown in Fig. \ref{figure1}(a). Here, we take the case $j\!=\!1$, $k\!=\!2$ as
an example for the nearest two atoms, and $j\!=\!1$, $k\!=\!10$ for the farthest two atoms. The
results reveal that, for infinite-range interaction such as $L\!=\!10^4d$, the interaction strengths
for the nearest and farthest two atoms are almost the same, as shown in the second row of Fig. \ref{figure4}.
While, when the localization lengths of the bound states decrease gradually, the difference of the interaction
strengths between the two cases becomes obvious. Specifically, for short-range interaction such as $L\!=\!d$,
the interaction strength between the nearest two atoms is approximately 0.37$V$, while the interaction between
the farthest two atoms becomes very weak and negligible, as shown in the last row of Fig. \ref{figure4}.
In other words, for finite localization length $L$, an atom can only exchange an excitation with proximal
atoms through its exponentially decaying envelope of the bound state, but not with the distant atoms. This
causes the phenomenon that the minimum resonance energy scales sub-linearly with the number $n$ of the
artificial atoms for finite localization length, as shown in Fig. \ref{figure3}(a).

\begin{figure}[tpb]    
\centering\includegraphics[width=8.6cm]{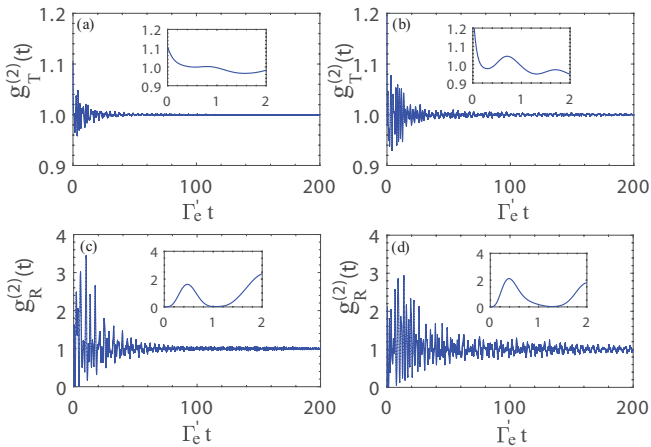} \caption{ Photon-photon correlation functions $g_{T}^{(2)}$ of the transmitted field and $g_{R}^{(2)}$ of the reflected field in the resonant case $\Delta=0$
when $n$ artificial atoms are equally spaced along the superconducting microwave photonic crystal.
Parameters: (a) and (c) $n=10$, (b) and (d) $n=20$; (a)-(d) $\mathcal {E}=0.0001\sqrt{\frac{\Gamma_{_{0}}}{2v_{g}}}$, $V=2.3\Gamma_{_{0}}$, $L\!=\!4.1d$, $k_{e}d\!=\!\pi/2$, $\Delta_{c}\!=\!0$, $\Omega_{c}\!=\!4\Gamma_{_{0}}$, $\Gamma_{e}'=9.5\times10^{-6}\Gamma_{_{0}}$, and $\Gamma_{s}'=9.5\times10^{-6}\Gamma_{_{0}}$. } \label{figure5}
\end{figure}

\subsection{Two-photon correlation} \label{correlation}

\begin{figure}[tpb]    
\centering\includegraphics[width=8.0cm]{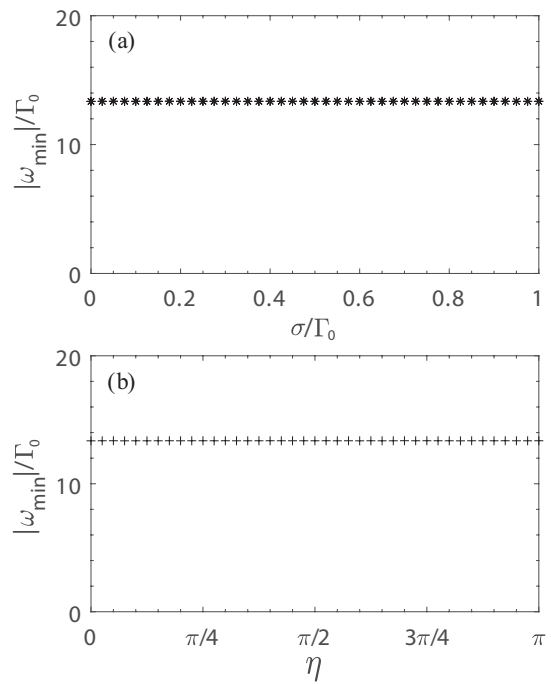} \caption{ The frequency $|\omega_{min}|$ calculated from exact numerics
versus the parameters (a) $\sigma$ and (b) $\eta$ when $n\!\!=\!\!10$ artificial atoms
are equally spaced along the superconducting microwave photonic crystal. We average
over 10 000 single-shot realizations for each $\sigma$ ($\eta$) with $\mathcal {E}=0.0001\sqrt{\frac{\Gamma_{_{0}}}{2v_{g}}}$,
$V=2.3\Gamma_{_{0}}$, $L\!=\!4.1d$, $k_{e}d\!=\!\pi/2$, $\Delta_{c}\!=\!0$,
$\Omega_{c}\!=\!4\Gamma_{_{0}}$, $\Gamma_{e}'=9.5\times10^{-6}\Gamma_{_{0}}$, and $\Gamma_{s}'=9.5\times10^{-6}\Gamma_{_{0}}$. } \label{figure6}
\end{figure}

The main feature of nonclassical light is that photons can be bunched or antibunched in time,
which is revealed via photon-photon correlation function $g^{(2)}(t)$ (second-order coherence). For
a steady state in our system, the photon-photon correlation function $g^{(2)}$ of the output field can be written as
\begin{eqnarray}     \label{eqa9}       
\text{g}_{\alpha}^{(2)}(\tau)\!\!=\!\!\frac{\langle\psi|a_{_{\alpha}}^{\dagger}(z)e^{iH\tau}a_{_{\alpha}}^{\dagger}(z)a_{_{\alpha}}(z)e^{-iH\tau}a_{_{\alpha}}(z)|\psi\rangle}{|\langle\psi|a_{_{\alpha}}^{\dagger}(z)a_{_{\alpha}}(z)|\psi\rangle|^{2}},
\end{eqnarray}
where $\alpha\!=\!T, R$, and $|\psi\rangle$ represents the steady-state wave vector.

Now, we turn to discuss the photon-photon correlation function $g^{(2)}(t)$ of the transmitted (reflected) field for two
choices of the number $n$ of the artificial atoms in the resonant case ($\Delta=0$), i.e., $n=10$ and $n=20$. Here, $n$
$\Delta$-type artificial atoms are equally spaced along the superconducting microwave photonic crystal. As shown in
Fig. \ref{figure5}, we observe oscillation between bunching and antibunching in both $g_{T}^{(2)}$ and $g_{R}^{(2)}$ for two
cases, i.e., $g^{(2)}$ oscillates around the uncorrelated value 1. Differently, for the transmitted field, initial
bunching ($g_{T}^{(2)}(0)>1$) exists, and initial antibunching ($g_{R}^{(2)}(0)=0$) is present for the reflected field.
By comparing the first and second rows of Fig. \ref{figure5}, we find that the amplitude of the oscillation in $g_{R}^{(2)}$ is stronger
than that in $g_{T}^{(2)}$ for both $n=10$ and $n=20$. The time delay of the oscillation in $g_{R}^{(2)}$ is longer than that
in $g_{T}^{(2)}$. Moreover, when we increase the number $n$ of the artificial atoms coupled to the photonic crystal, we observe
that the oscillation lasts longer in both $g_{R}^{(2)}$ and $g_{T}^{(2)}$. This can be concluded by comparing the first and second
columns of Fig. \ref{figure5}. The results mentioned above suggest that our system may offer an effective platform to explore the
nonclassical light in experiment.

\subsection{The effects of imperfections} \label{imperfection}

In above discussions, we have assumed that the coupling strengths between the photonic
crystal and artificial atoms are the same. While due to small experimental imperfections,
the coupling strengths are not exactly equal in practical condition \cite{NMSun2018}. Here,
we consider that the inhomogeneous broadening of the coupling strength is Gaussian with the
probability density $\rho(\delta)=\frac{1}{\sigma\sqrt{2\pi}}\exp({-\frac{\delta^{2}}{2\sigma^{2}}})$,
where $\delta$ is the difference value from the expected coupling strength $\Gamma_{_{0}}$, and $\sigma$ denotes the
standard deviation to measure the width of the inhomogeneous broadening.
As shown in Fig. \ref{figure6}(a), we give the dip frequency $|\omega_{min}|$ obtained
from exact numerics as a function of the parameter $\sigma$. The results show that the
frequency $|\omega_{min}|$ is robust to the inhomogeneous broadening of the coupling strength.
In fact, as shown in Figs. \ref{figure2}(a) and \ref{figure2}(b), the frequency $|\omega_{min}|$
originates from the long-range interaction term in Eq. (\ref{eqa2}). We conclude that, although
the coupling strengths between the photonic crystal and artificial atoms are not exactly equal
in experiment, our model is valid.

As mentioned above, the transitions $|e\rangle\leftrightarrow|s\rangle$ of the artificial atoms
are driven by external microwave fields.
In practical superconducting microwave photonic crystal systems, the phase of the external
driving field for each artificial atom may be different, which have been regarded to be
identical in above sections. Here, we assume that the external classical field driving the
$j$th artificial atom has a Rabi frequency $\Omega_{c}$ and a phase $\phi_{j}$, where
$\Omega_{c}$ is regarded to be identical for all artificial atoms and $\phi_{j}$ is randomly chosen from
[0, $\eta$] in our calculations. As shown in Fig. \ref{figure6}(b), we plot the dip frequency $|\omega_{min}|$
calculated from exact numerics as a function of the parameter $\eta$. Evidently, the dip
frequency $|\omega_{min}|$ is immune to the phase of the external driving field. In other
words, our model remains valid even the phases of the driving field for the artificial atoms
are different.

\begin{figure}[tpb]    
\centering\includegraphics[width=8.5cm]{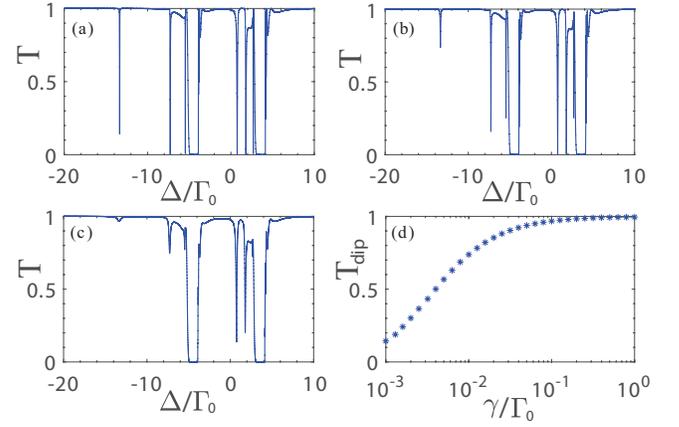} \caption{ The transmission spectra of the incident
field as a function of the detuning $\Delta/\Gamma_{0}$ for (a) $\gamma/\Gamma_{0}=0.001$, (b)
$\gamma/\Gamma_{0}=0.01$ and (c) $\gamma/\Gamma_{0}=0.1$ when $n=10$ artificial atoms are equally
spaced along the superconducting microwave photonic crystal. (d) The transmission $T_{dip}$ at the
minimum resonance frequency versus the dissipation rate $\gamma$ of the bound states.
Parameters: (a)-(d) $\mathcal {E}=0.0001\sqrt{\frac{\Gamma_{_{0}}}{2v_{g}}}$, $V=2.3\Gamma_{_{0}}$,
$L\!=\!4.1d$, $k_{e}d\!=\!\pi/2$, $\Delta_{c}\!=\!0$, $\Omega_{c}\!=\!4\Gamma_{_{0}}$, $\Gamma_{e}'=9.5\times10^{-6}\Gamma_{_{0}}$,
and $\Gamma_{s}'=9.5\times10^{-6}\Gamma_{_{0}}$. } \label{figure7}
\end{figure}

In an infinite superconducting microwave photonic crystal, the bound states can lead to permanent light trapping.
However, for a finite system, the dissipation of the bound states exists. Considering this dissipation, the long-rang interaction
Hamiltonian should be rewritten as
\begin{eqnarray}     \label{eqa10}       
H_{b}^{'}=-(V\!+\!\frac{i\gamma}{2}){{\sum\limits_{j, k}^n}}(-1)^{|z_{_{j}}-z_{_{k}}|/d}e^{-|z_{_{j}}-z_{_{k}}|/L}\sigma_{sg}^{j}\sigma_{gs}^{k},
\end{eqnarray}
where $\gamma$ denotes the dissipation rate of the bound states, assumed to be same for all the bound states.
For an infinite photonic crystal, $\gamma=0$. For an finite system, $\gamma$ is proportional to the overlap of
the bound states with the edges of the structure, i.e., $\gamma\sim e^{-l/L}$, where $l$ is the length of the system.
As shown in Fig. \ref{figure7}(a)-\ref{figure7}(c), we calculate the transmission spectra of the incident field for
three choices of the dissipation rate, i.e., $\gamma=0.001\Gamma_{0},0.01\Gamma_{0}$ and $0.1\Gamma_{0}$. We find that
the existence of the minimum-energy dip is sensitive to the dissipation rate of the bound states. For example, when we
change the rate $\gamma$ from $0.001\Gamma_{0}$ to $0.1\Gamma_{0}$, the dip almost disappears. In addition, to show this
phenomenon more clearly, we calculate the transmission $T_{dip}$ at the minimum resonance frequency as a function of the
dissipation rate $\gamma$, as shown in Fig. \ref{figure7}(d). The results reveal that, to observe the minimum-energy dip,
the dissipation of the bound states should be strongly suppressed. In fact, for a realistic superconducting photonic crystal \cite{YLiu2017,NMSun2018},
the dissipation rate is in the range of $\gamma\simeq[20\text{MHz}-120\text{MHz}]$.
Thus, $\gamma/\Gamma_{0}\simeq[6.0\times10^{-5}-3.6\times10^{-4}]$, with which we can
observe the minimum-energy dip clearly in the transmission spectrum.

\section{EXPERIMENTAL FEASIBILITY}   \label{expermental}

In this section, we discuss the experimental feasibility of our system. The phenomena and conclusions mentioned above may be
verified in recent experiments  \cite{YLiu2017,NMSun2018}. In their devices, a superconducting microwave photonic crystal can be implemented by periodically
alternating sections of low and high impedance coplanar waveguides. In the experiment reported by Sundaresan et al. \cite{NMSun2018},
two transmon qubits coupled to a superconducting microwave photonic crystal and widely tunable on-site and inter-bound state interactions
have been realized. In their device, the two coplanar waveguide sections are designed to be $d_{l}=7.8$ mm, $Z_{l}=124$ $\Omega$ and $d_{s}=1.2$ mm, $Z_{s}=25$ $\Omega$. Under such condition, the frequency of the second band edge is $\omega_{0}/2\pi=7.8$ GHz, and $v_{p}=1.248\times10^{8}$ m/s.
Each artificial atom is individually tunable in frequency by a local
flux bias line, and the atomic resonance frequency is adjusted to be $\omega_{sg}/2\pi=7.73$ GHz. Since $L=d\sqrt{\alpha/\delta}$ ($\alpha$ is the band curvature at the band edge), the localization length $L=4.1d$ adopted in our simulations can be obtained with $\alpha/2\pi=1.155$ GHz. In fact, longer-range interactions can be realized by reducing the detuning $\delta$ of the artificial atom from the band edge. For example, $L\simeq[3.7d-22.2d]$ has been reported by Liu et al. \cite{YLiu2017}.
In addition, the free space emission rate of the state $|s\rangle$ ($|e\rangle$) is $\Gamma_{s}^{'}/2\pi=0.5$ MHz ($\Gamma_{e}^{'}/2\pi=0.5$ MHz), and single-atom coupling strength to the second band is $g/2\pi=0.5275$ GHz.
Thus, the choices of the parameters in our numerical calculations may be experimentally feasible, except for the case $L=10^{4}d$.
In fact, this case ($L=10^{4}d$) is merely chosen as an idealized point of comparison.
Besides the transmon circuits,  a $\Delta$-type three-level artificial atom can also be realized by a flux-based superconducting circuit \cite{GuPR2017}.
Thus, to reach the ultrastrong-coupling regime, the transmon in their devices can be replaced by a flux qubit \cite{Forn2017,Yoshihara2017}.
Besides the method shown in the two experiments \cite{YLiu2017,NMSun2018}, there are other approaches to realize
superconducting microwave photonic crystals, such as lumped element circuits and Josephson
junction arrays.

\section{Conclusion}   \label{discussion}

In conclusion, we have studied the transport properties of a $\Delta$-type artificial
atomic chain coupled to an infinite superconducting microwave photonic crystal. In this work,
we take advantage of the superconducting quantum circuits, which provide widely tunable artificial
atoms with long coherence times \cite{Nori2011,houck2012}. Here, to reach the strong-coupling or ultrastrong-coupling
regimes, the artificial atomic chain can be realized by superconducting transmon qubits
\cite{YLiu2017,NMSun2018} or flux qubits \cite{Forn2017,Yoshihara2017,Liu2005,Jia2017,Kockum2019}.

In our system, we place the resonance frequency $\omega_{sg}$ of the
transition $|g\rangle\leftrightarrow|s\rangle$ inside the band gap, which gives rise to
the long-range coherent dipole-dipole interactions between the artificial atoms. To probe
the above mechanism in many-body regime, we utilize the coupling between the atomic transition
$|g\rangle\leftrightarrow|e\rangle$ and the second band modes of the superconducting microwave
photonic crystal. According to the phenomena concluded from exact numerics, we present a
simple model to estimate the dip frequency with known system parameters. The results
reveal that our model agrees well with exact numerics when the localization length is sufficiently
large. We also calculate the photon-photon correlation function of the output field, and observe oscillation between
bunching and antibunching for both transmitted and reflected fields. Moreover, we analyze the influence of a Gaussian inhomogeneous broadening of the coupling strength, different phases of the external driving field for the artificial atoms, and the dissipation of the bound states on the dip frequency, respectively. We show numerically that, the above conclusions still hold when the coupling strengths between the photonic crystal and artificial atoms are not exactly equal and the
phases of the driving field for the artificial atoms are different in practical condition. With the
dissipation rate of the bound states in a realistic superconducting photonic crystal, we can observe the minimum-energy dip clearly in the transmission spectrum. That is, with our model, one may acquire valuable system parameters by measuring the dip frequency in the transmission spectrum, which opens up a new avenue to explore the superconducting microwave photonic crystal systems.

\section*{ACKNOWLEDGMENTS}

We would like to thank W. Nie for helpful discussions.
This work is supported by the National Natural Science Foundation of China under Grants
No. 11175094 and No. 91221205, and the China Postdoctoral Science Foundation under Grant
No. 2017M620732. L.-C.K acknowledges support from the National Research Foundation and
Ministry of Education, Singapore. F.-G.D. is supported
by the National Natural Science Foundation of China under Grant No. 11674033.
G.-L.L. acknowledges support from the Center of Atomic and Molecular Nanosciences, Tsinghua
University, and Beijing Advanced Innovation Center for Future Chip (ICFC).


\begin{thebibliography}{64}



\bibitem{ShenOL2005} J. T. Shen and S. Fan, Opt. Lett \textbf{30}, 2001 (2005).


\bibitem{Shen2007PRL}   J. T. Shen and S. Fan,  Phys. Rev. Lett. \textbf{98}, 153003 (2007).


\bibitem{Shen2009pra}  J. T. Shen and S. Fan, Phys. Rev. A \textbf{79}, 023837 (2009).


\bibitem{Fan2010pra}  S. Fan, \c{S}. E. Kocaba\c{s}, and J. T. Shen, Phys. Rev. A \textbf{82}, 063821 (2010).


\bibitem{Zhou2013prl}  L. Zhou, L. P. Yang, Y. Li, and C. P. Sun,  Phys. Rev. Lett. \textbf{111}, 103604 (2013).



\bibitem{PLodahl2015rmp}  P. Lodahl, S. Mahmoodian, and S. Stobbe,  Rev. Mod. Phys. \textbf{87}, 347 (2015).


\bibitem{Liao2016}  Z. Liao, X. Zeng, H. Nha, and M. S. Zubairy,  Phys. Scr. \textbf{91}, 063004 (2016).


\bibitem{Diby2017rmp}  D. Roy, C. M. Wilson, and O. Firstenberg,  Rev. Mod. Phys. \textbf{89}, 021001 (2017).


\bibitem{Das2017PRL}    S. Das, V. E. Elfving, S. Faez, and A. S. S{\o}rensen, Phys. Rev. Lett. \textbf{118}, 140501 (2017).


\bibitem{Manzoni2017}   M. T. Manzoni, D. E. Chang, and J. S. Douglas, Nat. Commun. \textbf{8}, 1743 (2017).


\bibitem{Xia2018PRL}   K. Xia, F. Nori, and M. Xiao, Phys. Rev. Lett. \textbf{121}, 203602 (2018).



\bibitem{DEChang2018}  D. E. Chang, J. S. Douglas, A. Gonz\'{a}lez-Tudela, C.-L. Hung, and H. J. Kimble, Rev. Mod. Phys. \textbf{90}, 031002 (2018).



\bibitem{Shen2018pra}  Z. H. Chen, Y. Zhou, and J. T. Shen, Phys. Rev. A \textbf{98}, 053830 (2018).



\bibitem{AkimovNature2007}  A. V. Akimov, A. Mukherjee, C. L. Yu, D. E. Chang, A. S. Zibrov, P. R. Hemmer, H. Park, and M. D. Lukin, Nature (London) \textbf{450}, 402 (2007).


\bibitem{Chang2007nap}  D. E. Chang, A. S. S{\o}rensen, E. A. Demler, and M. D. Lukin, Nat. Phys. \textbf{3}, 807 (2007).



\bibitem{Tudela2011prl}   A. Gonzalez-Tudela, D. Martin-Cano, E. Moreno, L. Martin-Moreno, C. Tejedor, and F. J. Garcia-Vidal, Phys. Rev. Lett. \textbf{106}, 020501 (2011).



\bibitem{Akselrod2014}   G. M. Akselrod, C. Argyropoulos, T. B. Hoang, C. Cirac\`{\i}, C.
Fang, J. Huang, D. R. Smith, and M. H. Mikkelsen, Nature Photonics \textbf{8}, 835 (2014).


\bibitem{DayanScience2008}   B. Dayan, A. S. Parkins, T. Aoki, E. P. Ostby, K. J. Vahala, and H. J. Kimble, Science \textbf{319}, 1062 (2008).



\bibitem{Vetsch2010prl}  E. Vetsch, D. Reitz, G. Sagu\'{e}, R. Schmidt, S. T. Dawkins, and A. Rauschenbeutel, Phys. Rev. Lett. \textbf{104}, 203603 (2010).


\bibitem{RausPRL2011}  S. T. Dawkins, R. Mitsch, D. Reitz, E. Vetsch, and A. Rauschenbeutel, Phys. Rev. Lett.  \textbf{107}, 243601 (2011).


\bibitem{DReitz2013PRL}  D. Reitz, C. Sayrin, R. Mitsch, P. Schneeweiss, and A. Rauschenbeutel, Phys. Rev. Lett. \textbf{110}, 243603 (2013).


\bibitem{Petersen2014}  J. Petersen, J. Volz, and A. Rauschenbeutel, Science \textbf{346}, 67 (2014).


\bibitem{Kien2014}   F. Le Kien and A. Rauschenbeutel,  Phys. Rev. A \textbf{90}, 023805 (2014).


\bibitem{Mitsch2014}  R. Mitsch, C. Sayrin, B. Albrecht, P. Schneeweiss, and A. Rauschenbeutel, Nat. Commun. \textbf{5}, 5713 (2014).


\bibitem{GoutaudPRL2015}  B. Gouraud, D. Maxein, A. Nicolas, O. Morin, and J. Laurat, Phys. Rev Lett. \textbf{114}, 180503 (2015).


\bibitem{CSayrin2015}  C. Sayrin, C. Clausen, B. Albrecht, P. Schneeweiss, and A. Rauschenbeutel, Optica \textbf{2}, 353 (2015).



\bibitem{HLsorensen2016}  H. L. S{\o}rensen, J. B. B\'{e}guin, K. W. Kluge, I. Iakoupov, A. S. S{\o}rensen, J. H. M\"{u}ller, E. S. Polzik, and J. Appel, Phys. Rev. Lett. \textbf{117}, 133604 (2016).


\bibitem{Song2017AOP} G. Z. Song, Q. Liu, J. Qiu, G. J. Yang, F. Alzahrani, A. Hobiny, F. G. Deng, and M. Zhang, Ann. Phys. (NY) \textbf{387}, 152 (2017).


\bibitem{PSolano2017}  P. Solano, P. B. Blostein, F. K. Fatemi, L. A. Orozco, and S. L. Rolston, Nat. Commun. \textbf{8}, 1857 (2017).



\bibitem{Cheng2017pra}   M. T. Cheng, J. P. Xu, and G. S. Agarwal,  Phys. Rev. A \textbf{95}, 053807 (2017).


\bibitem{song2017pra}   G. Z. Song, E. Munro, W. Nie, F. G. Deng, G. J. Yang, and L. C. Kwek, Phys. Rev. A \textbf{96}, 043872 (2017).


\bibitem{Yanwb2018}    W. B. Yan, W. Y. Ni, J. Zhang, F. Y. Zhang, and H. Fan, Phys. Rev. A  \textbf{98}, 043852 (2018).


\bibitem{Yan2018}   C. H. Yan, Y. Li, H. D. Yuan, and L. F. Wei,  Phys. Rev. A \textbf{97}, 023821 (2018).


\bibitem{Cheng2018PRA}    D. C. Yang, M. T. Cheng, X. S. Ma, J. P. Xu, C. J. Zhu, and X. S. Huang, Phys. Rev. A \textbf{98}, 063809 (2018).


\bibitem{BabinecNat2010} T. M. Babinec, J. M. Hausmann, M. Khan, Y. Zhang, J. R. Maze, P. R. Hemmer, and M. Lon\v{c}ar,  Nat. Nanotechnol. \textbf{5}, 195 (2010).


\bibitem{ClaudonPhoton2010}   J. Claudon, J. Bleuse, N. S. Malik, M. Bazin, P. Jaffrennou, N. Gregersen, C. Sauvan, P. Lalanne, and J. M. G\'{e}rard,  Nat. Photon. \textbf{4}, 174 (2010).


\bibitem{Clevenson2015}   H. Clevenson, M. E. Trusheim, C. Teale, T. Schr\"{o}der, D. Braje, and D. Englund, Nat. Phys. \textbf{11}, 393 (2015).


\bibitem{Sipahigil2016}  A. Sipahigil, R. E. Evans, D. D. Sukachev, M. J. Burek, J. Borregaard, M. K. Bhaskar, C. T. Nguyen, J. L. Pacheco,
H. A. Atikian, C. Meuwly, R. M. Camacho, F. Jelezko, E. Bielejec, H. Park, M. Lon\v{c}ar, and M. D. Lukin, Science \textbf{354}, 847 (2016).


\bibitem{WallraffNature2004}   A. Wallraff, D. I. Schuster, A. Blais, L. Frunzio, R. S. Huang, J. Majer, S. Kumar,
S. M. Girvin, and R. J. Schoelkopf,  Nature (London) \textbf{431}, 162 (2004).


\bibitem{ShenPRL2005}  J. T. Shen and S. Fan, Phys. Rev. Lett. \textbf{95}, 213001 (2005).



\bibitem{Lzhou2008}  L. Zhou, Z. R. Gong, Y. X. Liu, C. P. Sun, and F. Nori,  Phys. Rev. Lett. \textbf{101}, 100501 (2008).


\bibitem{AstafievScience2010}  O. Astafiev, A. M. Zagoskin, A. A. Abdumalikov, Y. A. Pashkin,
T. Yamamoto, K. Inomata, Y. Nakamura, and J. S. Tsai,  Science \textbf{327}, 840 (2010).


\bibitem{LooSci2013}   A. F. van Loo, A. Fedorov, K. Lalumi\'{e}re, B. C. Sanders, A. Blais, and A. Wallraff,  Science \textbf{342}, 1494 (2013).


\bibitem{klalum2013}   K. Lalumi\`{e}re, B. C. Sanders, A. F. van Loo, A. Fedorov, A. Wallraff, and A. Blais,  Phys. Rev. A \textbf{88}, 043806 (2013).


\bibitem{GuPR2017}   X. Gu, A. F. Kockum, A. Miranowicz, Y. X. Liu, and F. Nori,  Phys. Rep. \textbf{718-719}, 1 (2017).


\bibitem{Kockum2018}   A. F. Kockum, G. Johansson, and F. Nori,  Phys. Rev. Lett. \textbf{120}, 140404 (2018).


\bibitem{MMirho2018}  M. Mirhosseini, E. Kim, V. S. Ferreira, M. Kalaee, A. Sipahigil, A. J. Keller, and O. Painter,  Nat. Commun. \textbf{9}, 3706 (2018).


\bibitem{PMHarr2018}   P. M. Harrington, M. Naghiloo, D. Tan, and K. W. Murch,  arXiv:1812.04205 (2018).


\bibitem{JDJoan2008book}  J. D. Joannopoulos, S. G. Johnson, J. N. Winn, and R. D. Meade, Photonic Crystals: Molding the Flow of Light,
2nd ed. (Princeton University Press, Princeton, 2008).


\bibitem{TudelaNAT2015}   A. Gonz\'{a}lez-Tudela, C. L. Hung, D. E. Chang, J. I. Cirac, and H. J. Kimble, Nat. Photonics \textbf{9}, 320 (2015).


\bibitem{JSDouglas2015}  J. S. Douglas, H. Habibian, C. L. Hung, A. V. Gorshkov, H. J. Kimble, and  D. E. Chang, Nat. Photonics \textbf{9}, 326 (2015).


\bibitem{jsDoug2016prx}  J. S. Douglas, T. Caneva, and D. E. Chang, Phys. Rev. X \textbf{6}, 031017 (2016).


\bibitem{EWAN2017}  E. Munro, L. C. Kwek, and D. E. Chang,  New J. Phys. \textbf{19}, 083018 (2017).


\bibitem{Cirac2018pra}  A. Gonz\'{a}lez-Tudela and J. I. Cirac, Phys. Rev. A \textbf{97}, 043831 (2018).


\bibitem{Song2018}  G. Z. Song, E. Munro, W. Nie, L. C. Kwek, F. G. Deng, and G. L. Long, Phys. Rev. A \textbf{98}, 023814 (2018).




\bibitem{GanL2015}   L. Gan and Z. Y. Li,  Sci. China-Phys. Mech. Astron. \textbf{58}, 114203 (2015).



\bibitem{SJohn1990prl}  S. John and J. Wang, Phys. Rev. Lett. \textbf{64}, 2418 (1990).


\bibitem{SJohn1990prb}   S. John and J. Wang,  Phys. Rev. B \textbf{43}, 12772 (1991).


\bibitem{SJohn1995prl}   S. John and T. Quang, Phys. Rev. Lett. \textbf{74}, 3419 (1995).




\bibitem{Rabl2016pra}  G. Calaj\'{o}, F. Ciccarello, D. E. Chang, and P. Rabl,  Phys. Rev. A \textbf{93}, 033833 (2016).



\bibitem{ShiT2016}  T. Shi, Y.-H. Wu, A. Gonz\'{a}lez-Tudela, and J. I. Cirac, Phys. Rev. X \textbf{6}, 021027 (2016).


\bibitem{Bello2018}  M. Bello, G. Platero, J. I. Cirac, and A. Gonz\'{a}lez-Tudela,  arXiv:1811.04390 (2018).


\bibitem{TLHansen2008}    T. Lund-Hansen, S. Stobbe, B. Julsgaard, H. Thyrrestrup, T. S\"{u}nner, M. Kamp, A. Forchel, and P. Lodahl, Phys. Rev. Lett. \textbf{101}, 113903 (2008).


\bibitem{ALaucht2012}  A. Laucht, S. P\"{u}tz, T. G\"{u}nthner, N. Hauke, R. Saive, S.
Fr\'{e}d\'{e}rick, M. Bichler, M.-C. Amann, A. W. Holleitner, M. Kaniber, and J. J. Finley, Phys. Rev. X \textbf{2}, 011014 (2012).


\bibitem{JDThompson2013}  J. D. Thompson, T. G. Tiecke, N. P. de Leon, J. Feist, A. V. Akimov,
M. Gullans, A. S. Zibrov, V. Vuleti\'{c}, and M. D. Lukin,  Science \textbf{340}, 1202 (2013).



\bibitem{AGobannatc2014}  A. Goban, C.-L. Hung, S.-P. Yu, J. D. Hood, J. A. Muniz, J. H. Lee,
M. J. Martin, A. C. McClung, K. S. Choi, D. E. Chang, O. Painter, and H. J. Kimble, Nat. Commun. \textbf{5}, 3808 (2014).


\bibitem{TGTiecke2014}  T. G. Tiecke, J. D. Thompson, N. P. de Leon, L. R. Liu,
V. Vuleti\'{c}, and M. D. Lukin, Nature (London) \textbf{508},  241 (2014).


\bibitem{MArcari2014prl}  M.  Arcari,  I.  S\"{o}llner,  A.  Javadi,  S.  Lindskov  Hansen,  S. Mahmoodian, J. Liu, H. Thyrrestrup, E. H. Lee, J. D. Song, S.  Stobbe,  and  P.  Lodahl, Phys.  Rev.  Lett.  \textbf{113},  093603  (2014).



\bibitem{SPYuapl2014}  S.-P. Yu, J. D. Hood, J. A. Muniz, M. J. Martin, R. Norte,
C.-L. Hung, S. M. Meenehan, J. D. Cohen, O. Painter, and H. J.  Kimble, Appl. Phys. Lett. \textbf{104}, 111103 (2014).


\bibitem{Tudela2015prl}   A. Gonz\'{a}lez-Tudela, V. Paulisch, D. E. Chang, H. J. Kimble, and J. I. Cirac, Phys. Rev. Lett. \textbf{115}, 163603 (2015).


\bibitem{ABYoung2015}   A. B. Young, A. C. T. Thijssen, D. M. Beggs, P. Androvitsaneas, L. Kuipers, J. G. Rarity, S. Hughes, and R. Oulton, Phys. Rev. Lett. \textbf{115}, 153901 (2015).



\bibitem{AGoban2015PRL}  A. Goban, C.-L. Hung, J. D. Hood, S.-P. Yu, J. A. Muniz, O. Painter, and H. J. Kimble, Phys. Rev. Lett. \textbf{115}, 063601 (2015).


\bibitem{Sollner2015}  I. S\"{o}llner, S. Mahmoodian, S. L. Hansen, L. Midolo, A. Javadi, G. Kir\v{s}ansk\.{e}, T. Pregnolato, H. El-Ella, E. H. Lee, J. D. Song, S. Stobbe, and P. Lodahl, Nat. Nanotechnol. \textbf{10}, 775 (2015).


\bibitem{Lefeber2015}  B. Le Feber, N. Rotenberg, and L. Kuipers,  Nat. Commun. \textbf{6}, 6695 (2015).


\bibitem{JDHood2016PNAS}  J. D. Hood, A. Goban, A. Asenjo-Garcia, M. Lu, S.-P. Yu, D. E. Chang, and H. J. Kimble, Proc. Natl. Acad. Sci. U.S.A. \textbf{113}, 10507 (2016).


\bibitem{Litao2018}   T. Li, A. Miranowicz, X. Hu, K. Xia, and F. Nori, Phys. Rev. A \textbf{97}, 062318 (2018).


\bibitem{Yu2018arxiv}   S. P. Yu, J. A. Muniz, C. L. Hung, and H. J. Kimble, arXiv:1812.08936 (2018).


\bibitem{Tudela2018arxiv}   A. Gonz\'{a}lez-Tudela and F. Galve, arXiv:1810.08155 (2018).


\bibitem{Burgersa2019}  A. P. Burgersa, L. S. Penga, J. A. Muniza, A. C. McClunga, M. J. Martina, and H. J. Kimble, Proc. Natl. Acad. Sci. U.S.A. \textbf{116}, 456 (2019).



\bibitem{YLiu2017}  Y. Liu and A. A. Houck, Nat. Phys. \textbf{13}, 48 (2017).


\bibitem{NMSun2018}  N. M. Sundaresan, R. Lundgren, G. Zhu, A. V. Gorshkov, and A. A. Houck, Phys. Rev. X \textbf{9}, 011021 (2019).




\bibitem{Liu2005}  Yu-xi Liu, J. Q. You, L. F. Wei, C. P. Sun, and F. Nori, Phys. Rev. Lett. \textbf{95}, 087001 (2005).


\bibitem{Jia2017}   W. Z. Jia, Y. W. Wang, and Yu-xi Liu, Phys. Rev. A \textbf{96}, 053832 (2017).


\bibitem{Chang2012}   D. E. Chang, L. Jiang, A. V. Gorshkov, and H. J. Kimble, New J. Phys. \textbf{14}, 063003 (2012).


\bibitem{Albrecht2017njp}  A. Albrecht, T. Caneva, and D. E. Chang, New J. Phys. \textbf{19}, 115002 (2017).



\bibitem{CanevaNJP2015}   T. Caneva, M. T. Manzoni, T. Shi, J. S. Douglas, J. I. Cirac, and D. E. Chang, New J. Phys. \textbf{17}, 113001 (2015).


\bibitem{Chang2011njp} D. E. Chang, A. H. Safavi-Naeini, M. Hafezi, and O. Painter, New J. Phys. \textbf{13}, 023003 (2011).



\bibitem{Forn2017}    P. Forn-D\'{\i}az, J. J. Garc\'{\i}a-Ripoll, B. Peropadre, J.-L. Orgiazzi, M. A. Yurtalan, R. Belyansky, C. M. Wilson, and A. Lupascu, Nat. Phys. \textbf{13}, 39 (2017).


\bibitem{Yoshihara2017}   F. Yoshihara, T. Fuse, S. Ashhab, K. Kakuyanagi, S. Saito, and K. Semba,  Nat. Phys. \textbf{13}, 44 (2017).


\bibitem{Nori2011}  I. Buluta, S. Ashhab, and F. Nori, Rep. Prog. Phys. \textbf{74}, 104401 (2011).


\bibitem{houck2012}   A. A. Houck, H. E. T\"{u}reci, and J. Koch,  Nat. Phys. \textbf{8}, 292 (2012).




\bibitem{Kockum2019}   A. F. Kockum, A. Miranowicz, S. De Liberato, S. Savasta, and F. Nori,  Nat. Rev. Phys. \textbf{1}, 19 (2019).




\end{thebibliography}
\end{document}